# PRINCIPLE OF PROCESSES DISCERNIBILITY

V.A. Etkin

The principle of classification of macroprocesses is offered, which allows to avoid many methodological mistakes of modern natural sciences. The content and the nearest consequences of this principle is open and its conformity of a classical line Physics developments is shown

# ПРИНЦИП РАЗЛИЧИМОСТИ ПРОЦЕССОВ

Д.т.н., проф. В.Эткин

Предложен принцип классификации макропроцессов, который позволяет избежать многих методологических ошибок современного естествознания. Раскрывается содержание и ближайшие следствия этого принципа и показывается соответствие его классической линии развития физики.

**Введение.** Известно, что в пространственно неоднородных системах одни и те же изменения состояния (например, повышение температуры) могут быть вызваны как внешним теплообменом, так и появлением внутренних источников тепла вследствие трения, химических реакций, высокочастотного нагрева, перемагничивания и т.п. Аналогичным образом объемную деформацию системы можно вызвать не только совершением работы сжатия, но и самопроизвольным расширением в пустоту. Точно так же одинакового изменение состава системы можно добиться как диффузией подходящих веществ через границы системы, так и химическими реакциями в ней. Следовательно, в любой теории неравновесных систем процессы должны классифицироваться не зависимо от того, чем вызваны те или иные изменения состояния – внешним энергообменом или внутренними (в том числе релаксационными) процессами.

Это положение является одним из исходных принципов термодинамики неравновесных процессов переноса и преобразования энергии [1]. В ней процессы классифицируются лишь по их *последствиям*, т.е. по особым, феноменологически отличимым и несводимым к другим изменениям состояния, которые эти процессы вызывают[1]. Такие процессы целесообразно для краткости называть *феноменологически различимыми*. Именно этого принципа придерживалась и классическая термодинамика, различая изохорный и изобарный, изотермический и адиабатический процессы [2]. Принят он и в других фундаментальных дисциплинах. В частности, в классической механике этому принципу классификации соответствуют процессы перемещения и ускорения тел, а также поступательного и вращательного движения. В электротехнике таким же образом различают электризацию и поляризацию тел[2]; поляризацию и намагничивание; электропроводность и потокосцепление. В химии на этом основано различение гомогенных и гетерогенных реакций. Тот же принцип соблюдается при классификации многочисленных эффектов на стыках различных научных дисциплин [3].

Это положение целесообразно назвать для удобства ссылки *принципом различимости процессов*, сформулировав его следующим образом: «*Существуют процессы, вызывающие специфические, феноменологически различимые и не сводимые к другим изменения состояния системы*» [4].

---

[1] В этом её отличие от физической кинетики, которая классифицирует процессы по причинам, их вызывающим, различая, в частности, концентрационную диффузию, термодиффузию и бародиффузию; а также от теории теплообмена, которая различает процессы по механизму переноса энергии и делит его на кондуктивный, конвективный и лучистый.

[2] В первом случае вносится заряд одного знака, что вызывает отклонение лепестков электроскопа; во втором – разделение разноименных зарядов, что вызывает появление разности потенциалов.

Представляет интерес выяснить, какие последствия вызывает принятие этого подчас само собой разумеющегося положения.

**1. Различимость процессов как основа отыскания их координат.** Признание возможности принять за основу классификации процессов принцип их различимости предъявляет определенные требования к выбору их «координаты», т.е. *физической величины, изменение которой является необходимым и достаточным признаком протекания этого процесса*. Эти требования состоят в выборе в качестве координаты процесса только такого параметра, который не изменяется при одновременном протекании в тех же элементах пространства других, так же независимых процессов [5]. Именно этим была обусловлена проблема нахождения в классической термодинамике координаты процесса теплообмена, поскольку ею мог стать только такой параметр, который в отличие от температуры или объема не изменялся бы в адиабатических процессах, т.е. воздействиях любого иного рода, кроме теплообмена. Как известно, такой параметр был найден основоположником термодинамики Р. Клаузиусом в 1865 году и назван им (не вполне удачно)[1] энтропией. Этот параметр позволял различать теплоту и работу, однако такое разделение всех видов энергообмена системы породило целый ряд трудностей, не преодоленных полностью до настоящего времени. Дело в том, что термин «теплота» до сих пор употребляется в двух смыслах: как функция *состояния* (называемая кратко «теплотой тела») и как функция *процесса*, (кратко называемая «*теплотой процесса*») и служащая количественной мерой теплообмена. Эта двойственность понимания теплоты возникла исторически в связи с ее рассмотрением как хаотической формы движения (в одном ряду с такими явлениями, как свет, звук, электричество, магнетизм) и сохранилась вопреки многочисленным дискуссиям. Понимание теплоты как формы энергии закрепилось в понятии теплоемкости системы. Оно проявилось также и в теории теплообмена, ибо обмениваться можно только тем, чем располагает система. В неравновесных системах такое понимание теплоты диктуется целым рядом тепловых эффектов, вызванных диссипацией (трением, высокочастотным или индукционным нагревом, химическими превращениями), т.е. не связанных с теплообменом. Лишь в равновесных системах, где такого рода тепловые эффекты отсутствуют, теплота становится количественной мерой процесса теплообмена. Поэтому в равновесной термодинамике она трактуется как «*энергия в состоянии перехода*», т.е. как «*то, что подводится через границы системы*».

Однако деление энергообмена на теплоту $Q$ и работу $W$ стало невозможным после введения в рассмотрение ещё одного вида энергообмена – массообмена, который обусловлен переносом вещества через границы системы и не сводим ни к теплоте, ни к работе. Поскольку этот процесс сопровождается переносом как внутренней тепловой энергии, так и работой деформации системы, «на границе, где имеет место диффузия вещества, понятия «теплота» и «работа» утрачивают свой смысл» [6].

Еще более очевидным стала невозможность нахождения координат независимых процессов энергообмена с переходом к изучению процессов изменения состава системы, которые могут быть вызваны как диффузией различных веществ через границы системы, так и химическими реакциями внутри её самой. Таким образом, при рассмотрении неравновесных систем стало принципиально невозможным отыскание координат независимых процессов *энергообмена*. Возникла необходимость перехода к иной классификации процессов и к нахождению их координат вне зависимости от того, чем вызваны изменения состояния системы – внешним энергообменом (как в равновесных системах) или внутренними необратимыми (релаксационными) процессами. Здесь-то и становится необходим принцип различимости процессов.

Покажем это на примере «теплопроцесса»[2], под которым понимается изменение внутренней тепловой энергии системы $dU_т$ как благодаря подводу тепла $dQ$ извне, так и вследствие выделения тепла диссипации $dQ^д$ в самой системе[3]:

---

[1] Термин «энтропия» (внутреннее превращение) характеризует другой процесс, связанный с диссипацией.
[2] Понятие, введенное К. Путиловым в 1971 [7].

$$dU_\text{т} = đQ + đQ^\text{д}. \qquad (1)$$

Именно эту величину Р.Клаузиус назвал «полной теплотой тела», определив её как сумму п]лученной извне теплоты и работы «дисгрегации» (диссипативного характера) [8]. В этом случае частное от деления полного дифференциала одного параметра состояния $U_\text{т}$ на другой параметр (абсолютную температуру $T$) заведомо является полным дифференциалом, так что существование энтропии $S$ как параметра состояния и координаты теплопроцесса $S$ не требует сложных обоснований [7]:

$$dS = dU_\text{т}/T = đQ/T + đQ^\text{д}/T = d_eS + d_iS, \qquad (2)$$

где $d_eS$, $d_iS$ – изменения энтропии, вызванные соответственно внешним теплообменом и внутренними источниками тепла диссипации. Именно в таком виде записал впервые баланс энтропии в необратимых процессах нобелевский лауреат И. Пригожин [9].

Аналогичным образом объем системы $V$ является координатой процесса объемной деформации не зависимо от того, чем она вызвана – работой расширения или расширением в пустоту без совершения работы. Точно так же числа молей $k$-х веществ $N_\text{k}$ могут служить координатами процесса изменения состава системы, не зависимо от того, чем они обусловлены – диффузией этих веществ через границы системы или внутренними химическими превращениями.

Как видим, классификация процессов по принципу их различимости не только устраняет необходимость идеализации процессов, выраженной терминами «обратимый», «равновесный», «квазистатический», но и позволяет обобщить понятие координаты процесса на случай их изменения в отсутствие внешнего энергообмена.

Не менее важным следствием классификации процессов по принципу их различимости является отказ от деления всех видов энергообмена на теплоту $Q$ и работу $W$. Неизбежность переноса внутренней тепловой энергии $U_\text{т}$ при переносе вещества через границы системы лишает возможности определить теплоту лишь как «одну из форм энергообмена» [7]. Осознание того, что подвод тепла извне вызывает в системе те же изменения состояния, что и работа диссипативного характера, вынуждает рассматривать теплообмен как своего рода «микроработу» против хаотически направленных межмолекулярных сил. Такого же рода «микроработа» совершается и при вводе в систему массы или любого $k$-го вещества, а с ним – и некоторого заряда. Как и работа всестороннего сжатия, эти виды работ не связаны с преодолением результирующей каких-либо сил из-за хаотчности молекулярного движения в системе. Поэтому все эти виды работ, изменяющих только внутреннюю энергию системы, целесообразно отнести к категории «неупорядочных», отличая их тем самым от «упорядоченных» работ, связанных с изменением внешней энергии системы. Принципиальное различие «упорядоченных» и «неупорядочных» работ проявляется в том, что первые из них являются количественными мерами процесса превращения энергии из одной формы в другую, а вторые – количественными мерами процесса её переноса в одной и той же форме [5]. Отсюда – понимание того, что истинная «линия водораздела» проходит не между теплотой и работой, а между двумя этими категориями работ.

Немаловажным следствием предложенной классификации работ является возможность различать упорядоченную (внешнюю) $E$ и неупорядоченную (внутреннюю) энергию системы $U$. Это позволяет определить полную энергию системы как её способность совершать любую (упорядоченную и неупорядоченную) работу и тем самым вернуть ей близкий к изначальному простой и ясный физический смысл. Это очень близко к данному Максвеллом определению энергии как «суммы всех действий, которые может оказать система на окружающие ее тела». Последнее особенно важно в связи с известным высказыванием нобелевского лауреата Р.Фейнмана о том, что «современная физика не знает, что такое энергия».

---

[3]) Предложенный С. Ньюменом в 1875 знак неполного дифференциала $đ$ означает, что в данном случае имеется в виду не изменение какой-либо величины $Q$ или $W$, а их элементарное количество.

Эвристическая ценность принципа различимости процессов подтверждена всей многовековой историей успешного развития естествознания. Именно с изучения свойств того или иного объекта и выяснения специфики протекающих в нем процессов начинается его экспериментальное исследование. Эта специфика лежит и в основе классификации явлений, что подтверждается самим фактом «ветвления» единого древа науки по мере углубления знаний.

**2. Принцип адекватности описания объекта исследования.** Использование энергодинамической классификации процессов по их последствиям позволяет обосновать принципиально важную теорему, согласно которой *число независимых координат, определяющих состояние любой (равновесной или неравновесной) энергодинамической системы, т.е. число её степеней свободы, равно числу феноменологически различимых процессов, протекающих в ней.*

Это положение легко доказывается «от противного». Поскольку под процессом понимается изменение свойств системы, выраженных параметрами состояния, то при их протекании с необходимостью изменяется хотя бы один из них. Предположим, что при протекании какого-либо феноменологически различимого процесса с необходимостью изменяются несколько координат состояния. Тогда, очевидно, эти координаты не будут независимыми, что противоречит исходной посылке. Предположим теперь, что какая-либо из координат состояния изменяется с необходимостью при протекании нескольких процессов. Тогда, очевидно, эти процессы не будут феноменологически различимыми, поскольку они вызывают одни и те же изменения свойств системы. Остается заключить, что *любому* (равновесному или неравновесному, квазистатическому или нестатическому) *феноменологически различимому процессу соответствует единственная независимая координата состояния.* Такие координаты – в общем случае величины экстенсивные, поскольку каждая из них в отсутствие других степеней свободы определяет энергию системы – величину также экстенсивную [5].

Доказанное положение определяет *необходимые и достаточные* условия адекватного (достаточно полного) задания состояния той или иной системы. Поэтому для удобства ссылки его целесообразно назвать «*принципом адекватности*» описания состояния. Этот принцип позволяет избежать как «недоопределения», так и «переопределения» системы[1], что является главным источником методологических ошибок и парадоксов современной термодинамики [5]. Далеко не очевидно, например, «недоопределение» состояния континуума, к которому приводит принятие гипотезы локального равновесия. Эта гипотеза исключает необходимость присутствия в основном уравнении неравновесной термодинамики градиентов температур, давлений и других обобщенных потенциалов (т.е. термодинамических сил) на том основании, что в элементах объема предполагается равновесие. Тем самым исключаются внутренние силы (напряжения), приводящие к изменению внутреннего состояния таких элементов. С другой стороны, не очевидно и «переопределение» континуальной среды, вызванное приписыванием ей бесконечного числа степеней свободы вопреки конечному числу протекающих в каждом её элементе макропроцессов.

Одним из важных следствий принципа адекватности является возможность отличить термодинамические параметры системы от нетермодинамических. Согласно ему, к первым следует отнести лишь те, величины, которые являются количественной мерой того или иного свойства системы. Это означает, что такие свойства системы, как цвет, вкус, запах и т.п. не могут считаться термодинамическими параметрами состояния. Это относится, в частности, и к «обонятельным», «осязательным» и т.п. степеням свободы, произвольно введенным в термодинамику А. Вейником (1968) [10].

---

[1] Т.е. попыток описать состояние системы недостаточным или избыточным числом координат.

Характерно, что методологические ошибки, связанные с недоопределением или переопределением исследуемых систем, обнаружить труднее всего. Тем тяжелее их последствия.

**3. Решение проблемы термодинамических неравенств.** Следуя доказанной выше теореме, легче выделить в объекте исследования феноменологически различимые процессы. Это можно сделать с помощью всего арсенала экспериментальных средств, поскольку различить процессы можно не только по причинам, их порождающим, и не столько по специфическими условиями их протекания, но и в особенности по их последствиям. Так мы устанавливаем необходимое и достаточное число независимых координат состояния, однозначно определяющих её полную энергию Э. В пространственно неоднородных средах эти координаты подразделяются на равновесные и неравновесные. К первым относятся уже известные параметры состояния $\Theta_i$ (масса системы $M$, её объем $V$, числа молей $k$-х веществ $N_k$, энтропия $S$, заряд $З$ и т.п.). Ко вторым относятся параметры, характеризующие векторные процессы релаксации, протекающие в таких системах и приводящие к выравниванию температур, давлений, химических, электрических и тому подобных потенциалов в различных частях системы. Эти процессы, как показано в [5,11], приводят к перераспределению указанных выше величин $\Theta_i$ по объему системы и смещению их центра на величину $\Delta \mathbf{r}_i$ от его положения в однородной (внутренне равновесной) системе. В результате этого возникают некоторые «моменты распределения» $\mathbf{Z}_i = \Theta_i \Delta \mathbf{r}_i$ «энергоносителей» $\Theta_i$, число которых в общем случае равно числу степеней свободы той же системы в её однородном состоянии. Таким образом, полная энергия неравновесной системы Э оказывается функцией удвоенного числа экстенсивных переменных Э $=$Э$(\Theta_i, \mathbf{R}_i)$, а её полный диференциал принимает вид:

$$d\text{Э} \equiv \Sigma_i \psi_i d\Theta_i - \Sigma_i \mathbf{F}_i \cdot d\mathbf{r}_i. \qquad (3)$$

где $\psi_i \equiv (\partial E/\partial \Theta_i)$ – обобщенные потенциалы (от латинского *potentia* – сила); $\mathbf{F}_i \equiv -(\partial E/\partial \mathbf{r}_i)$ – силы в их обычном (ньютоновском) понимании [12]. Введение в термодинамику изначально чуждого ей понятия силы $\mathbf{F}_i$ делает естественным последующее введение в нее времени $t$, скорости $\mathbf{v}_i = d\mathbf{r}_i/dt$ и производительности $N_i = \mathbf{F}_i \cdot \mathbf{v}_i$ реальных процессов.

Выражение (3) носит характер тождества и потому справедливо при любых значениях входящих в него величин, не зависимо от того, учитывают ли они вклад диссипативных процессов или нет. Поэтому оно применимо *к любым процессам* (как обратимым, так и необратимым). Благодаря этому достигается решение известной проблемы термодинамических неравенств, состоящей в переходе уравнения (3) в неравенство в случае необратимых процессов. Это происходит вследствие того, что в неравновесных системах элементарная неупорядоченная работа $đW_i^{\text{н}} \neq \Sigma_i \psi_i d\Theta_i$, а упорядоченная работа $đW_i^{\text{e}} \neq \mathbf{F}_i \cdot d\mathbf{r}_i$ ввиду возможности самопроизвольного изменения координат $\Theta_i$ и $\mathbf{r}_i$ [13]. Это означает, что члены тождества (3) уже не определяют величину каждой из упорядоченных или неупорядоченных работ, поскольку они включают в себя часть работы диссипативного характера.. Такова цена, которую приходится платить за исключение неравенства в выражении (3).

**4. Неразличимость процессов в современной физике**. Еще в 1632 году в книге "Диалог о двух главнейших системах мира – птолемеевой и коперниковой" Галилей отметил как факт, что если на движущемся прямолинейно и равномерно корабле отпустить камень с мачты, то он падает так же, как и на неподвижном корабле – к подножию мачты. В частности, в трюме корабля, плывущего равномерно и прямолинейно, никакими экспериментами невозможно обнаружить его движение относительно водной среды и суши. Это положение, получившее в механике название «*принципа относительности Галилея*», утверждало, что равномерное и прямолинейное движение одной системы материальных тел относительно другой совершенно не сказывается на ходе механических процессов, происходящих внутри этих материальных систем. И.Ньютон положил этот принцип в основание его 1-го закона (постулата), сформулировав его следующим образом: «всякое тело продолжает удерживаться в состоянии покоя или равномерного и прямолинейного движения, пока и поскольку оно не понуждается приложенными силами изменить это состояние" [14]. Отсюда следовало, что никакими механическими опытами, производимыми внутри замкнутой механической системы, нельзя установить, покоится ли данное тело или движется равномерно и прямолинейно.

А. Пуанкаре в 1895 году распространил этот принцип на электромагнитные явления, назвав его *постулатом относительности.* Согласно ему, не только механическими, но и электромагнитными опытами, производимыми внутри произвольной системы отсчета, нельзя установить различие между состояниями покоя и равномерного прямолинейного движения. Отсюда следовало, что физические законы должны формулироваться таким образом, чтобы покой и равномерное прямолинейное движение системы были *неразличимы* [15].

А.Эйнштейн в 1905 году распространил постулат относительности на все явления природы и положил его в основание специальной теории относительности. Вскоре он же сформулировал принцип локальной неразличимости сил тяготения и сил инерции, назвав его принципом эквивалентности инерционной и гравитационной масс и положив его в основание общей теории относительности (ОТО). Затем к нему присоединился принцип неразличимости ускоренного и вращательного движений, который распространил неразличимость динамических эффектов ускорения и тяготения на неинерциальные системы отсчета [16]. Так принцип неразличимости покоя и равномерного прямолинейного движения (и инерциальных систем отсчета) стал основным исходным принципом теоретического построения всей физики и научного исследования. В электродинамике это выразилась в принципе неразличимости электронов в металле; в физике элементарных частиц – в принципе неразличимости тождественных частиц; в КЭД – в неразличимости вещества и поля; в единой теории поля – в утверждении о возможности слияния воедино (при определенных условиях – до полной неразличимости) трёх из четырех известных видов взаимодействия. И все это сделано на основе экстраполяции принципа Галилея, справедливого лишь для инерциальных систем. В результате известная идея Лейбница об отсутствии в природе двух совершенно тождественных вещей (его аналог принципа различимости) была заменена в современной теоретической физике принципами неопределенности, запрета и неразличимости, что сделало понимание физических процессов необязательным и в значительной мере иллюзорным. Постулирование этих принципов породило в конечном счете неразличение истины и заблуждений. В этой связи отказ от постулирования и экстраполяции принципа неразличимости означает возвращение физики на классический путь развития.

## Литература